\documentclass[12pt]{iopart}

\usepackage{graphicx}

\begin{document}

\title[STM of NiMnSb(001): an {\it ab-initio} study]{Scanning
  tunneling microscopy of surfaces of half-metals: an {\it ab-initio}
  study on NiMnSb(001)}

\author{M Le\v{z}ai\'c, Ph Mavropoulos, G Bihlmayer, and S Bl\"ugel}

\address{Institut f\"ur Festk\"orperforschung, Forschungszentrum J\"ulich, D-52425
J\"ulich, Germany}

\ead{m.lezaic@fz-juelich.de,ph.mavropoulos@fz-juelich.de}

\begin{abstract}
We present a first-principles study of the unreconstructed (001)
surfaces of the half-metallic ferromagnet NiMnSb. Both terminations
(MnSb and Ni) are considered. We find that half-metallicity is lost at
the surfaces. After a discussion of the geometric relaxations and the
spin-polarized surface band structure, we focus on topography images
which are expected to be found with spin-polarized scanning tunneling
microscopy. For the MnSb-terminated surface we find that only the Sb
atoms are visible, reflecting a geometric buckling caused by
relaxations. For the Ni-terminated surface we find a strong contrast
between the images of forward and reverse tip-sample-bias of 0.5 eV,
as well as a stripe-like image for reverse bias. We interpret these
findings in terms of highly directional surface states which are
formed in the spin-down gap region.
\end{abstract}




\maketitle


\section{Introduction\label{sec:1}}

Half-metallic Heusler alloys have attracted considerable attention
lately, because of their potential applications to spintronics. In
particular, the desirable half-metallic character (100\% spin
polarization at the Fermi level $E_F$), which renders them candidates
for efficient spin-dependent transport, has been predicted by numerous
calculations for many half- and full-Heusler
alloys~\cite{groot,Springer,Galanakis2002a,Galanakis2002b}. The
surfaces of Heusler alloys have also been examined, in the hope of
probing the half-metallic property there. It seems, though, that the
spin polarization is strongly reduced, or even reversed, at surfaces,
and that the half-metallic gap is lost: computational
results~\cite{Jenkins01,GalanakisSurf001,Jenkins04,Lezaic05Surf}
suggest that this is due to surface states which are formed by broken
bonds; spin-polarized photoemission spectroscopy does also not show
half-metallicity at the surfaces~\cite{Bona85}.

In a previous study~\cite{Lezaic05Surf} we presented first-principles
calculations on the NiMnSb(001) and (111) surfaces, showing how the
surface states appear within the gap and close it. We also described
in detail how these states can develop to interface states
\cite{Galanakis05} and how these interface states affect the
spin-dependent transport properties, in particular when half-metals
are used as leads in tunnel junctions~\cite{Mavropoulos05}. In this
paper we extend our work to the study of spin-dependent scanning
tunneling microscopy (SP-STM) images of the NiMnSb(001) surfaces. Our
focus is on the possibility to distinguish the two possible (001)
terminations (Ni and MnSb), and on the form of the topography pictures
for charge density, magnetization density, and spin-resolved density.

In section~\ref{sec:2} we briefly describe the method of
calculation. Our results are shown and discussed in
section~\ref{sec:3}, after a summary of our previous study of the
electronic structure~\cite{Lezaic05Surf}. We conclude with a summary
in section~\ref{sec:4}.

\section{Method of calculation\label{sec:2}}
The calculations were performed within the generalized gradient
approximation (GGA) \cite{PBE} of density functional theory. We used
the full-potential linearized augmented planewave (FLAPW) method in
film geometry \cite{Wimmer}, as implemented in the {\tt FLEUR} code
\cite{FLEUR}. For the calculations, a planewave cutoff $K_{\rm max}$
of 3.6~a.u.$^{-1}$ was used. The charge density and the wavefunctions
within the muffin-tin spheres
($R_{MT}\mathrm{(Ni,Mn)}=2.3~\mathrm{a.u.}=1.22$~\AA\ ;
$R_{MT}\mathrm{(Sb)}=2.4~\mathrm{a.u.}=1.27$~\AA) were expanded in
lattice harmonics with angular momentum $l\leq 8$. We modeled the
(001) surfaces using a slab of 9 atomic layers of NiMnSb. The
two-dimensional Brillouin zone was sampled with 64 special {\bf
k}-points in the irreducible wedge. All the calculations were
performed at the theoretically determined equilibrium in-plane lattice
constant of NiMnSb (5.915~\AA), which is within 0.2\% in agreement
with the experimental value \cite{Otto}. The atomic structure was
optimized by calculation of the forces on the three topmost layers of
the film.  Possible reconstructions were not considered.

\section{Results and discussion\label{sec:3}}

\subsection{Unrelaxed and relaxed surface structure}

\begin{figure}
\begin{center}
\includegraphics[angle=270,width=0.7\linewidth]{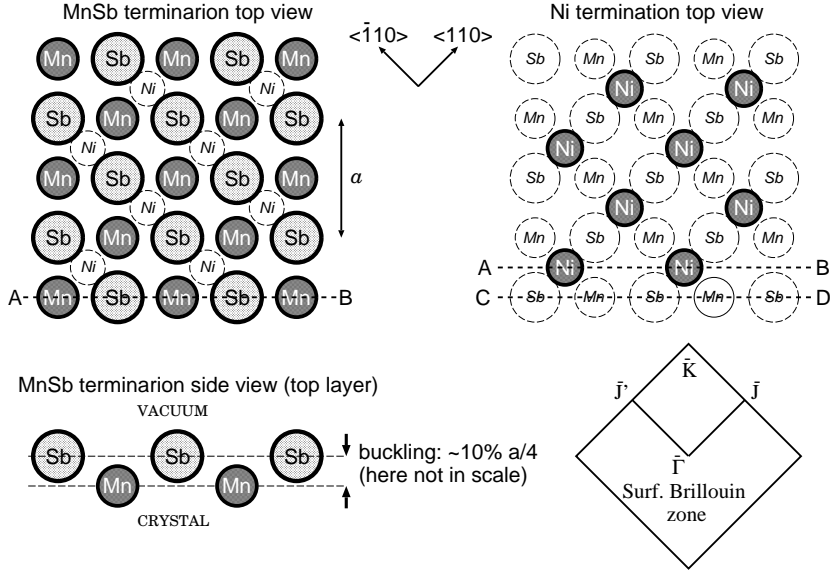}
\caption{Schematic top view of the MnSb-terminated (left) and the
  Ni-terminated (right) NiMnSb(001) surfaces. The sub-surface atomic
  positions are also shown (thin dashed circles) to demonstrate the
  $C_{2v}$ symmetry. By $a$ we denote the fcc lattice constant. The
  lines AB and CD were used for the STM line scans (see also
  Figure~\protect{\ref{fig:stm1}}). In absence of geometric
  relaxations the interlayer distance is $a/4$. The buckling of the
  topmost MnSb layer is also shown (not to scale). On the lower right
  the surface Brillouin zone is shown and the points $\bar{\Gamma}$,
  $\bar{J}$, $\bar{J}'$, and $\bar{K}$, appearing in the band
  structures of Figure~\protect{\ref{fig:bands}}, are
  indicated.\label{fig:struc}}
\end{center}
\end{figure}

NiMnSb crystallizes in the L2$_1$ structure, consisting of three
inter-penetrating fcc sublattices. Viewed as a succession of atomic
planes in the $\langle 001\rangle$ direction, the structure consists
of alternating Ni and MnSb layers, with the layer occupancy and
geometry repeating every fourth layer. The interlayer distance is
$a$/4 (with $a$ being the bulk lattice constant). Thus, chemically,
the (001) surface has two different possible terminations: one with Mn
and Sb at the surface, and one with Ni (see
Figure~\ref{fig:struc}). Geometrically, the layer occupancy is
repeated every fourth layer, since every second layer the atoms
exchange positions (Mn with Sb and Ni with the empty site; thus the
top view of every second layer is equivalent but rotated by
$90^\circ$). The surface symmetry group is $C_{2v}$. On the question
of surface reconstruction, we point out that Ristoiu and
collaborators~\cite{Ristoiu00} found an unreconstructed surface by low
energy electron diffraction experiments, and that Jenkins and
King~\cite{Jenkins01} report that the unreconstructed MnSb-terminated
surface is energetically lower than the reconstructed one.

In the calculations for the structural relaxation of the surfaces we
allowed in each case for the three top layers to relax. For the Ni
terminated surface we find that the distance $d$ to the subsurface
layer is reduced by 10\%, while there is no appreciable buckling or
relaxation of the subsurface MnSb layer. In the case of the MnSb
terminated surface, we find that for the Sb atoms $d$ is expanded by
7.3\%, while for the Mn atoms $d$ is contracted by 3.5\%. This gives
an overall buckling of 10.8\% of the interlayer distance $d=a/4$. This
is in reasonable agreement with the buckling determined by Jenkins and
King~\cite{Jenkins01}, and, as we see below, is important for the STM
images of the MnSb-terminated surface.

\subsection{Surface electronic and magnetic structure}

\begin{figure}
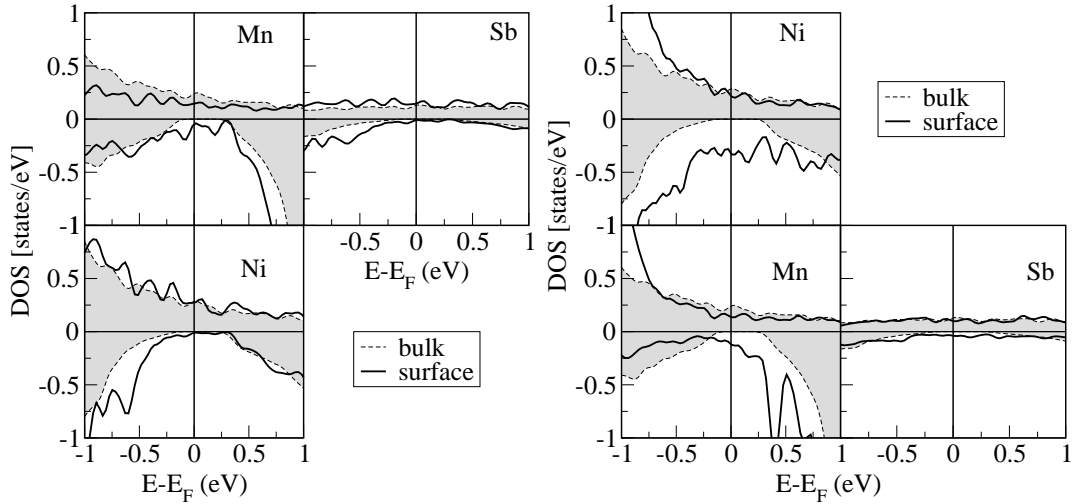

\begin{center}
\includegraphics[width=7cm]{DOS001MnSb.eps}
\includegraphics[width=7cm]{DOS001Ni.eps}
\caption{Atom-resolved DOS at the surface (upper panels) and
  subsurface (lower panels) layers for MnSb-terminated (left) and
  Ni-terminated (right) NiMnSb(001) surfaces.  The results of the
  relaxed surfaces are indicated by solid. Gray-shaded regions
  represent the bulk DOS.\label{fig:dos}}
\end{center}
\end{figure}

The densities of states (DOS) of the NiMnSb(001) surface and
sub-surface layers close to the Fermi level are shown in
Figure~\ref{fig:dos} together with the bulk DOS for
comparison. Clearly, the spin-down gap which is present in the bulk
closes at the surface. The surface states responsible for this loss of
half-metallicity come about due to the broken translational symmetry
at the surface. As pointed out in reference~\cite{Galanakis2002a}, the
spin-down gap at $E_F$ originates from the hybridization of the Mn $d$
orbitals with the Ni $d$ orbitals. Since this hybridization ceases at
the surface, the gap collapses. For a detailed study on the origin and
nature of these surface states we refer to the works of Jenkins and
King~\cite{Jenkins01} and Le\v{z}ai\'c {\it et
al.}~\cite{Lezaic05Surf}. We will discuss them further in connection
to the STM images in the next subsection.

\begin{figure}
\begin{center}
\includegraphics*[angle=270,width=14cm]{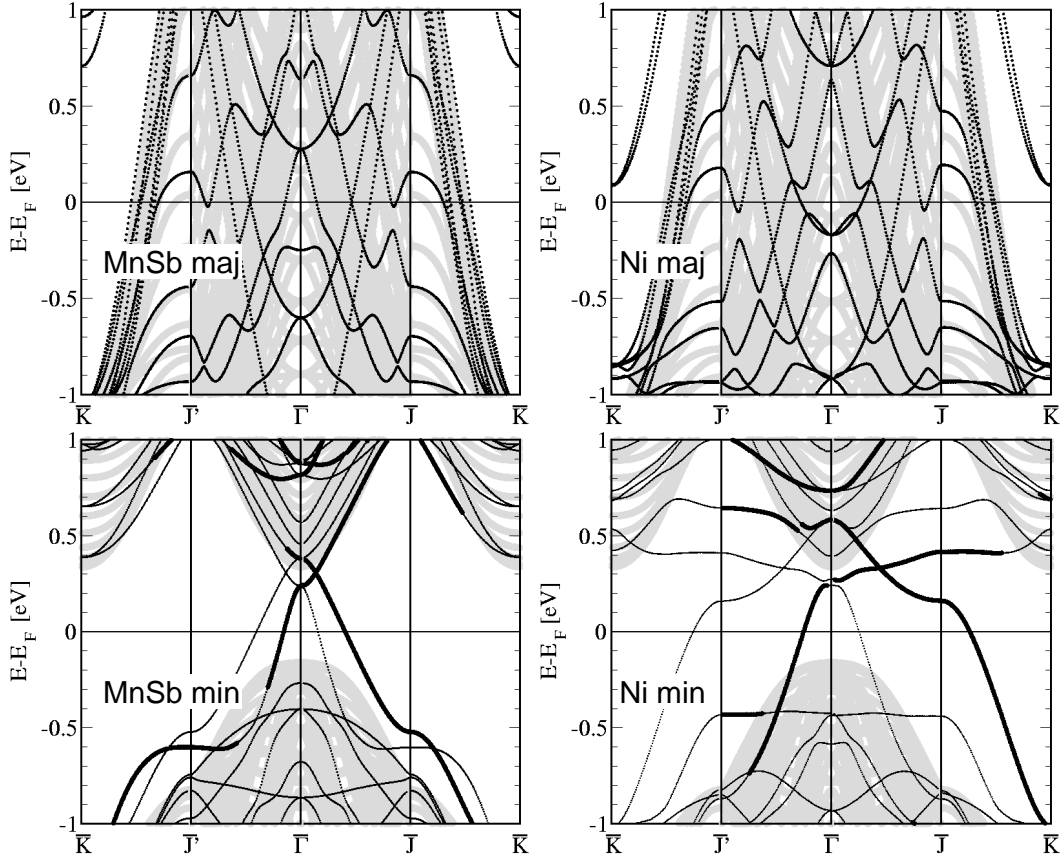}
\caption{Majority (upper panels) and minority-spin (lower panels)
surface band structure for the MnSb (left) and Ni (right) terminated
(001) surfaces. Gray regions indicate the surface-projected bulk band
structure. Thin lines indicate the result for the relaxed
surface. Thick lines indicate surface states on one of the of the two
equivalent surfaces of the film (we considered that a surface state
should have more than 50\% of its weight located at the first two
surface layers). For the notation of the points in the surface
Brillouin zone see Figure~\protect{\ref{fig:struc}}.\label{fig:bands}}
\end{center}
\end{figure}

The surface band structure of NiMnSb(001) is presented in
Figure~\ref{fig:bands}. Because the calculation involves a finite
slab, the surface bands come in pairs, one band for each slab
surface. We identify the bands of a single surface as those with more
than 50\% weight within the first two surface layers; this results in
the thickly marked bands of Figure~\ref{fig:bands}. 

\subsection{STM images}

For the STM topography images we use the model of Tersoff and
Hamann~\cite{Tersoff83}, according to which the tunneling current $I$
from or into the tip above a lateral surface point $(x,y)$ and at a
vertical distance $z$ from the surface at a bias voltage $U$ is
proportional to the integrated density of states $N(x,y,z;eU)$ at the
position of the tip,
\begin{equation}
I\sim N(x,y,z;eU)=\int_{E_F}^{E_F+eU} n(x,y,z;E)\, dE,
\end{equation}
for $eU>0$ (and the same but integrated from $E_F-|eU|$ to $E_F$ for
$eU<0$), with $n(x,y,z;E)$ being the local DOS of the sample at the
position of the tip apex.  We chose $eU=\pm 0.5$~eV and vertical
distances of $z=2.6$~\AA\ and 4~\AA\ (taken from the muffin-tin
boundary of the surface atoms; see section~\ref{sec:2}). Our results,
showing the integrated DOS, were qualitatively the same for both
distances, and below we show the images calculated for $z=4$~\AA\ from
the surface layer.

\begin{figure}
\begin{center}
\includegraphics[angle=270,width=10cm]{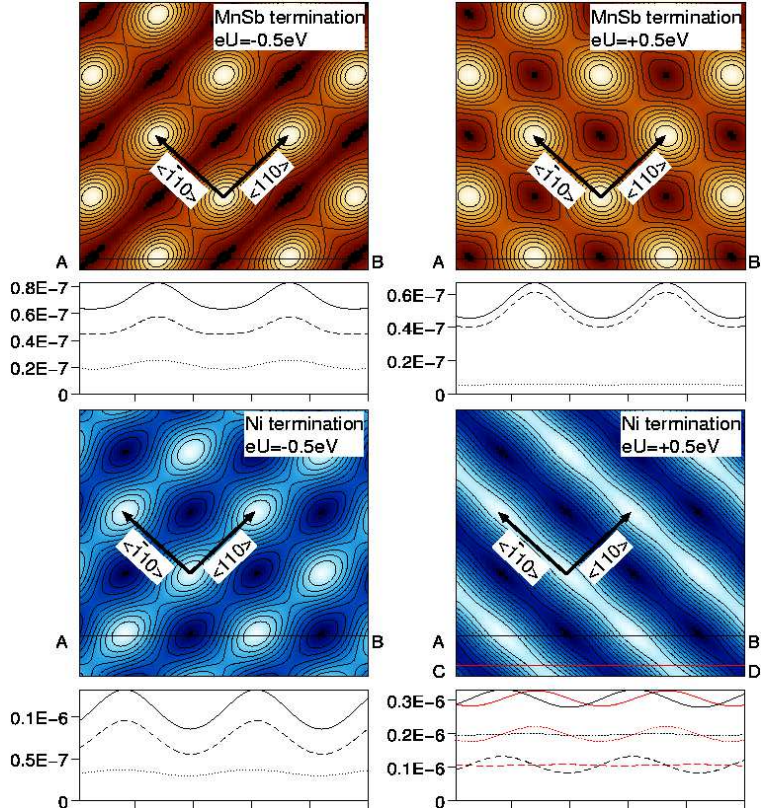}
\caption{(color online) STM topography of the MnSb-terminated surface
  (top panels) and of the Ni-terminated surface (bottom panels) at
  $z=4$~\AA\ for an area of $13\times 13$~\AA\ and for $eU=\pm
  0.5$~eV. Arrows indicate the surface Bravais vectors along the
  $\langle\bar{1}10\rangle$ and $\langle 110 \rangle$ directions, with
  origin and endpoints at Sb atoms for the MnSb case and at Ni atoms
  for the Ni terminated case.  A line scan along the line AB is also
  shown below each image. In the line scans, the dashed line
  corresponds to spin up, the dotted line to spin down, and the solid
  line to the total signal. For the case of Ni termination with $eU=+
  0.5$~eV, two line scans are shown: one along AB (black, passing over
  the atom positions) and one along CD (red, passing over the
  interatomic bridge position), in order to demonstrate the change of
  the corrugation from the spin-up to the spin-down
  states.\label{fig:stm1}}
\end{center}
\end{figure}

The STM topography for the MnSb-terminated surface is shown in
Figure~\ref{fig:stm1} (upper panels). The first striking effect is
that only the Sb atoms are visible, while above the Mn atoms the
tunneling current shows local minima. This is mainly caused by the
aforementioned 10.8\% buckling of the surface atoms, which renders the
retracted Mn atoms invisible to the STM tip. Thus, it is not possible
to distinguish the MnSb-termination by its higher atomic density
compared to the Ni-termination: the extra atoms are ``hidden''.

Nevertheless, a second striking effect is revealed when one compares
the topography for both signs of the voltage $U$. While for the MnSb
termination $N(x,y,z;eU)$ is comparable in magnitude for both bias
voltages $eU=\pm 0.5$~eV, for the Ni termination there is a difference
between $N(x,y,z;+0.5\mathrm{eV})$ and $N(x,y,z;-0.5\mathrm{eV})$ of a
factor three (the corrugation, however, is of the same
magnitude). This is demonstrated in Figure~\ref{fig:stm1} (bottom
panels), where we have plotted line scans of $N(x,y,z;eU)$ along the
$x$ direction. At the same time, the topographic image at $+0.5$~eV
shows stripes in the $\langle \bar{1}10\rangle$ direction, with the
positions of the atoms barely visible, together with a corrugation in
the $\langle 110\rangle$ direction. The image at $-0.5$~eV, on the
contrary, shows comparatively little directionality, with more
symmetric patterns around the Ni atoms.

Given this effect, the nature of the surface can be readily found in
an STM experiment by the strong contrast between forward- and
reverse-bias currents and images at the Ni-terminated surface, as
opposed to the comparable forward- and reverse-bias currents at the
MnSb-termination.

In order to explain this contrast, we observe that, for the
Ni-termination, surface states appear in the $\bar{\Gamma}-\bar{J}$
direction between 0.1~eV and 0.5~eV above $E_F$ (see
Figure~\ref{fig:bands}).  At this point we recall that most of the
tunneling current results from states around the $\bar{\Gamma}$-point,
but the topography images (the corrugation) are frequently determined
by states in the vicinity of high-symmetry points at the Brillouin
zone boundary~\cite{Heinze}. This is consistent with the observation
that the directionality of the surface states in $k$-space is
reflected in the stripy topographic pattern; the corrugation is in the
$\langle 110\rangle$ direction, as a result of the states close to the
surface Brillouin zone boundary $\bar{J}$. From the line scan
(Figure~\ref{fig:stm1}) it is also evident that the main contribution
to the STM intensity comes from spin-down electrons. The corrugation,
however, comes from both spin directions in an alternating pattern, as
can be seen from the difference between the line scans along AB and
CD.  The former one (AB) passes over the atoms, where the spin-up
states contribute to the expected corrugation.  The latter line scan
(CD) passes over the interatomic bridge positions, where the spin-down
$d$ orbitals, showing a directional preference, form those
one-dimensional surface states which are dispersive along the $\langle
\bar{1}10\rangle$ direction (represented by $\bar{\Gamma}-\bar{J}'$ in
Figure~\ref{fig:bands}) and non-dispersive along the $\langle
110\rangle$ direction (represented by $\bar{\Gamma}-\bar{J}$ in
Figure~\ref{fig:bands}). 

We interpret the dispersive form of these surface bands as
follows. Due to the subsurface atoms, the (001) surface has only a
twofold $C_{2v}$ symmetry with natural coordinate axes along the
$\langle 110\rangle$ direction (we call this the $\alpha$ axis) and
along the $\langle \bar{1}10\rangle$ direction (we call this the
$\beta$ axis). Then the $d_{\beta z}$ orbitals at the surface have an
overlap along the $\beta$ axis, resulting in dispersive bands, but no
overlap along the $\alpha$ axis, resulting in non-dispersive bands.
Therefore in the spin-down surface band structure of the Ni-terminated
surface (Figure~\ref{fig:bands}) we see, at the $\bar{\Gamma}$-point
and at about 0.2~eV, one dispersive band departing towards the
$\bar{\Gamma}-\bar{J}'$ direction and one non-dispersive band
departing towards the $\bar{\Gamma}-\bar{J}$ direction. The overlap of
the $d_{\beta z}$ orbitals is mediated by $d$-$d$ hybridization with
the underlying Mn atoms. On the contrary, the $d_{\alpha z}$ orbitals
are oriented towards the underlying Sb atoms, so that a corresponding
hybridization cannot take place (the $d_{\alpha z}$ states are
orthogonal to the $p_z$ states of the Sb neighbors). Therefore, we
obtain the second band at $\bar{\Gamma}$, at 0.5~eV, which is
non-dispersive in both directions. (For a discussion of
one-dimensional surface states in the MnSb-terminated surface, see
reference~\cite{Jenkins01}.)

In order to cross-check our explanation on the origin of the striped
pattern, we produced topographic images for bias voltages of $\pm
0.1$~V. We expect that the striped pattern and the high spin-down DOS
will disappear for the image of $+0.1$~eV, since the narrow spin-down
surface states are present only above $E_F+0.1$~eV. The images (not
shown here) verify our hypothesis, showing now fourfold symmetry, with
the density dominated by spin-up states; furthermore the intensity for
$-0.1$~eV appears twice as strong as for $+0.1$~eV, {\it i.e.}, the
situation is reversed compared to the 0.5~V case.

In view of the above results we conclude that STM topography can
distinguish the Ni-terminated from the MnSb-terminated (001) surface
by the contrast between forward- and reverse-bias images at 0.5~V, as
well as by the striped, corrugated pattern which appears for
$eU=0.5$~eV.

We conclude the discussion with a comment on spin-polarized STM images
for the Ni-terminated surface at $eU=+0.5$~eV. From the line scans it
is evident that the spin-up and spin-down corrugations are
complementary. This means that, in an {\it ideal} SP-STM experiment
the spin-up contrast will be stronger above the atoms (in accordance
with general wisdom for STM on metallic surfaces), while the spin-down
contrast will be stronger above the bridge positions (contrary to
general wisdom). The stripy features will disappear. We verified this
by plotting the corresponding two-dimensional STM images (not shown
here). In a real SP-STM experiment, of course, the difference between
the two spins will be captured to the extent of the spin-filtering
quality of the spin-polarized tip. 

\section{Summary and outlook\label{sec:4}}

We have investigated the NiMnSb(001) unreconstructed surfaces by
first-principles calculations. Emphasis was given on the STM
topographic images and their relation to the surface geometry and
electronic structure.

There are two possible terminations: a MnSb termination and a Ni
termination. In both cases we find that half-metallicity, which is
present in the bulk, is lost at the surface. This is attributed to the
breaking of $d$-$d$ hybridization bonds (between the Mn and Ni states)
which are actually responsible for the gap in the bulk. The surface
states which are thus formed are mainly of $d$ character.

For the MnSb-terminated surface, relaxation of the atomic positions
results in a buckling of about 10.8\% with respect to the interlayer
distance, with the Sb atoms protruding outwards. This buckling is
significant for the STM image, because it renders the Mn atoms
invisible to the STM tip. From previous work~\cite{Jenkins01} on the
MnSb surface, surface reconstruction is not expected.

For the Ni-terminated surface the STM images at $eU=+0.5$~eV show a
peculiar stripe-like corrugation. This partly originates from spin-up
states, which give a contrast above the Ni atoms, and partly from the
spin-down states, which give a contrast above the bridge
positions. The latter effect can be traced back to the form of the
spin-down $d$-like surface bands. In particular, the surface has
twofold symmetry, which is inherited by the surface band structure
because of the bonding via the underlying Mn atoms. As a result, the
spin-down surface states are dispersive in the direction
$\bar{\Gamma}-\bar{J}'$ and non-dispersive in $\bar{\Gamma}-\bar{J}$.
In addition, we find that the current is dominated by spin-down
states. These features disappear at bias voltages lower than
$eU=+0.1$~eV, which corresponds to the bottom of the narrow
$\bar{\Gamma}-\bar{J}$ spin-down surface bands; below this voltage,
the spin-down current and corrugation become unimportant.

As an outlook, we comment on the possibility of spin-polarized STM in
the case that a surface would be half-metallic. Such surfaces have
been predicted to exist in some half-metallic systems, including
transition-metal pnictides and chalcogenides in the zinc-blende
structure~\cite{GalanakisZB} or even the Mn-rich, Mn-terminated
Co$_2$MnSi(001) surface~\cite{Hashemifar05}. In such cases no
spin-down current should exist. However, since the tip itself is
normally not half-metallic, the spin-up (with respect to the sample)
current is present for both orientations of the tip magnetization
(parallel and antiparallel to the sample). Thus one should expect a
strong asymmetry by reversing the tip magnetization, but not an effect
of absolute current blockade. Additionally, there should be a drastic
increase of the current when the increased bias reaches the valence
band or the conduction band of the spin-down channel.

\section*{Acknowledgements}
We would like to thank Stefan Heinze for a critical reading of the
manuscript.

\end{document}